\documentclass{article}

\usepackage{euscript}

\usepackage{latexsym}
\usepackage{amsmath, amssymb,graphics}
\usepackage[all]{xy}

\usepackage{fullpage}

\numberwithin{equation}{section}

\renewcommand{\(}{\begin{equation*}}
\renewcommand{\)}{\end{equation*}}
\newcommand{\bea}{\begin{eqnarray*}}
\newcommand{\eea}{\end{eqnarray*}}
\newcommand{\R}{{\mathbb R}}
\newcommand{\C}{{\mathbb C}}
\newcommand{\Z}{{\mathbb Z}}

\newcommand{\cW}{\ensuremath{\mathcal W}}

\newcommand{\beq}{\begin{equation}}
\newcommand{\eeq}{\end{equation}}

\numberwithin{equation}{section}

\renewcommand{\(}{\begin{equation}}
\renewcommand{\)}{\end{equation}}

\newcommand{\ZZ}{{\mathbb Z}}

\def\R{{\mathbb R}}
\def\Z{{\mathbb Z}}

\def\C{{\mathbb C}}

\def\1{{\bf 1}}

\def\<{\langle}
\def\>{\rangle}

\numberwithin{equation}{section}

\renewcommand{\(}{\begin{equation}}
\renewcommand{\)}{\end{equation}}

\begin{document}


%
\begin{titlepage}


%

\begin{center}
{\Large\bf 
Twisted topological structures related to M-branes} 
\end{center}
\vspace{1em}

\begin{center}
Hisham Sati 
\footnote{e-mail: {\tt
hsati@math.umd.edu}}
\end{center}

\begin{center}
Department of Mathematics\\
University of Maryland\\
College Park, MD 20742 
\end{center}

\vspace{0.5cm}
\begin{abstract}
\noindent

Studying the M-branes leads us naturally to new structures
that we call Membrane-, Membrane${}^c$-, String${}^{K(\Z, 3)}$-
and Fivebrane${}^{K(\Z,4)}$-structures, which we
show can also have twisted counterparts. 
We study some of their basic properties,
highlight analogies with structures associated with 
lower levels of the Whitehead tower of the orthogonal 
group, and demonstrate the relations to M-branes.

\end{abstract}

\end{titlepage}

\tableofcontents

\section{Introduction}
This is a continuation of 
our study of geometric and topological structures related to 
M-branes in M-theory, and is the third paper in a series. In the first one
\cite{tcu} we outlined several ideas to be developed in later papers. 
In the second \cite{II} we considered
twisted String structures 
\cite{Wa} and introduced twisted String${}^c$ structures 
associated with 
the M2-brane and the M5-brane. This builds on 
\cite{SSS3} where twisted String structures are considered in relation to 
the flux quantization condition in M-theory \cite{Flux} and the
Green-Schwarz anomaly cancellation in heterotic string theory.
In addition,  the paper \cite{SSS3} considered
 the dual picture, where the anomalies 
associated with the M5-brane and the dual of the Green-Schwarz 
anomaly cancellation lead to twisted Fivebrane structures, 
a twisted version of the Fivebrane structure introduced in 
\cite{SSS1} and studied in \cite{SSS2}.

\vspace{3mm}
The obstructions for the String and Fivebrane cases are 
essentially the first and the second Pontrjagin classes, respectively. 
In this paper we consider, in addition, obstructions coming from the 
Stiefel-Whitney classes. The seventh integral Stiefel-Whitney class
$W_7$ of spacetime appears as a condition for the partition function 
in dimensionally reduced M-theory to type IIA to be well-defined
upon summing over torsion \cite{DMW}. This was taken in 
\cite{KS1} to provide an elliptic cohomology refinement of the 
K-theoretic partition function. In addition, such a construction 
led naturally to a condition on the mod 2 Steifel-Whitney class
$w_4$. This has a natural interpretation in relation to the flux
quantization of the C-field \cite{tcu}. 
In this paper we provide an interpretation of these 
structures in terms of bundle constructions and  also
consider twists for such structures. 
 The motivation comes from M-branes in M-theory, and so 
  we find such structures in relation to worldvolumes,
normal bundles of embedding to spacetime, and in 
target spacetime 
itself.

\vspace{3mm}
We define the following structures

\begin{enumerate}

\item {\it (Twisted) Membrane structures}. 
These are structures whose obstructions 
are degree four analogs of those of twisted Spin structures.
Instead of having $w_2 + b=0\in H^2\Z_2$, we have
the condition $w_4 + \alpha=0\in H^4\Z_2$, where $\alpha$ is a degree
four $\Z_2$ class. This shows up in the quantization condition for the 
 C-field and hence in relation to the 
partition function of the M2-brane, and also 
in the normal bundle of the M5-brane. We study this in section
 \ref{t M2}.
 
\item {\it (Twisted) Membrane${}^c$ structures}. 
A twisted membrane${}^c$ structure is a degree five analog of a twisted
Spin${}^c$ structure, where instead of having the Freed-Witten condition 
$W_3 + H_3=0\in H^3\Z$ \cite{FW}, interpreted as the obstruction to having
a twisted Spin${}^c$ structure \cite{Wa} \cite{Do}, we have 
$W_5 + H_5=0 \in H^5\Z$. Such a structure is also related to the C-field. 
We consider this in section \ref{t M2 c}.

\item {\it (Twisted) String${}^{K(\Z,3)}$ structures}. 
These are analogs of Spin${}^c$
structures in the sense that obstructions to their existence are given by an odd
degree Steenrod operation on a cohomology class.
Recall that the Spin${}^c$ condition is obtained by applying 
the 
Bockstein $\beta$ on the 
Stiefel-Whitney class $w_2$ giving $W_3$. Now $W_7$ is similarly obtained via 
a Steenrod square applied to a characteristic class, namely 
$Sq^3$ acting on $\lambda=\frac{1}{2}p_1$.
So we call such a structure a String${}^{K(\Z, 3)}$ structure.
We also consider the twist of such a structure and relate it to 
physical situations discussed in \cite{7} and \cite{tcu}.
This is the subject of section \ref{String KZ3}.

\item {\it (Twisted) Fivebrane${}^{K(\Z,4)}$ structures}. 
 String${}^c$ structure introduced in \cite{CHZ}
are String structures corresponding to Spin${}^c$ 
bundles.  Such structures are shown in \cite{II}
 to be related to twisted String structures, 
and a twisted version was provided. 
We might call this a twisted String${}^{K(\Z,2)}$ in
our present formalism.
In degree eight, we find that we replace $K(\Z,2)$ by 
$K(\Z,4)$, the first Chern class of the line bundle corresponding to 
the Spin${}^c$ structure with the characteristic class $a$ of an 
$E_8$ bundle, 
\footnote{Note that $E_8 \sim K(\Z,4)$ in our range of dimensions.}
and the String condition with the Fivebrane 
condition. Furthermore, this can be twisted by a degree eight
cocycle. 
 The result is what we call 
a twisted Fivebrane${}^{K(\Z,4)}$ structure, which we show is 
related to the dual C-field in M-theory, and hence to the M5-brane.
This will be discussed in section \ref{Fivebrane KZ4}.

\end{enumerate}
The above structures admit restrictions to the boundary, 
which occur and are relevant in all cases, but we illustrate only for the 
case of twisted membrane structures, with the others 
deduced directly from that case. This is relevant for 
M-theory in the presence of a boundary, and for the 
M2-brane and the M5-brane, both in the presence of boundaries, and in 
treating them essentially as boundaries.

%
%
%
%
%
%
%
%
%

\vspace{3mm}
As the conditions appearing in quantization conditions and 
anomaly cancellation conditions are given only up to further
denominators by 
such structures, we characterize the `mismatch' from the 
point of view of cohomology operations in section \ref{Sec cong}. 
This should be viewed as 
complentary  to the interpretations in \cite{SSS3} in terms of 
bundles encoding the extra congruences. 
In doing so we naturally highlight the importance of torsion.  
We find that the one-loop term in M-theory
\cite{DLM} \cite{VW} can be succinctly described using 
what we call {\it String characteristic classes}. Some of the relations 
between the various structures introduced in this paper are 
indicated in the sections where such structures are defined. 
Further connections are made in section \ref{relate}.

\section{Review of setting and basic notions}
\label{rev}
The goal of this section is to provide some motivation, setting, as well as 
tools and definitions 
which will be needed in the following sections.

\paragraph{Twisted Spin structure and type I fields.}
In the presence of a $B$-field $B$,
type I D-branes can be wrapped on a submanifold 
$D$ of spacetime only if \cite{W}
\(
w_2(D) + [B|_D]=0 \in H^2(M;\Z_2)\;.
\label{ko}
\)
A geometric interpretation of such 2-torsion B-fields (also called
t'Hooft classes) as
the holonomy of connections for real bundle gerbes
is given in \cite{MMS}, where 
both finite dimensional and infinite dimensional geometric realizations 
are given. 
In \cite{Wa} the condition  \eqref{ko} is interpreted as twisted Spin structure.

\paragraph{Twisted Spin${}^c$ structures and the Ramond-Ramond fields.}
In the presence of a Neveu-Schwarz (NS)  $B$-field, type II D-branes can
be wrapped  on a submanifold $D$ of spacetime only if \cite{FW}
\(
W_3(D) + [H|_D] =0 \in H^3(M;\Z)\;.
\label{FW}
\)
A geometric interpretation in terms of bundle gerbes
is given in \cite{BCMMS}. 
This condition  \eqref{FW} is not sufficient 
and further obstructions arise  at primes higher than 2 
\cite{ES2}. 

\vspace{3mm}
Generalizations of the above structures to higher levels in 
the Whitehead tower of the homotopy groups of the 
orthogonal group are also relevant. In \cite{Wa}
the notion of twisted String structure was defined. 
This was refined in \cite{SSS3} to the 
differential case, where also the physical applications
to the C-field and to the Green-Schwarz anomaly formula
are discussed. Furthermore, the notion of twisted
Fivebrane structure was introduced in \cite{SSS3}, with
the dual $C$-field and the dual Green-Schwarz 
anomaly formula viewed as (essentially) 
obstructions to such a structure. 

\paragraph{Twisted String structure and the $C$-field.}
The $C$-field, via its field strength $G_4$, 
in M-theory on a Spin eleven-manifold $Y$
satisfies the quantization condition \cite{Flux}
\(
[G_4] - \frac{\lambda}{2}= a \in H^4(Y;\Z)\;,
\label{flux}
\)
where $\lambda=\frac{1}{2}p_1$ is half the first Pontrjagin class of $Y$
(see \eqref{Q1}) and 
$a$ is the degree four characteristic class of 
an $E_8$ bundle over $Y$. We will sometimes
omit the notation $[x]$ for a cohomology class, and use 
just $x$ instead. 
\footnote{
This 
should not cause confusion since we will not deal with 
differential form representatives in this paper.}
The corresponding geometric structure is a 
twisted String structure \cite{SSS3}, in the sense of 
\cite{Wa}. 

\paragraph{Twisted Fivebrane structures and the dual $C$-field.}
The dual $C$-field, via its field strength $G_8$,
in M-theory on $Y$ satisfies the quantization condition \cite{DFM}
\(
\Theta:=[G_8]=\frac{1}{48} p_2 + \beta\;,
\label{dual flux}
\)
where $\beta$ is a degree eight class. This is interpreted 
as a twisted Fivebrane structure, defined in \cite{SSS3}. The untwisted case,
in which  $\beta$ is not present corresponding to $\frac{p_2}{6}=0$
was introduced in \cite{SSS1} \cite{SSS2}. 

\vspace{3mm}
The factors in the denominators of the Pontrjagin classes in 
\eqref{flux} and \eqref{dual flux} are interpreted in terms of 
twisted String and Fivebrane structure, respectively, which 
are modified in the appropriate sense \cite{SSS3}.
This will be studied further in this paper, in section
\ref{Sec cong}, from the point of view of cohomology operations. 
We will mostly deal with torsion in cohomology.

\paragraph{Description of the higher Stiefel-Whitney classes.}
Since there are flat manifolds $M$ with nonzero 
Stiefel-Whitney classes $w_k(TM)$ \cite{ASz}, 
one cannot hope for an analog of Chern-Weil theory for 
Stiefel-Whitney classes. However, one can hope for 
a \v Cech description. The cases $k=1,2 $ are well-known (see \cite{LM}).
An explicit \v Cech cocycle representing the $k$th 
Stiefel-Whitney class of an $n$-dimensional vector bundle over a manifold 
$M$ is given in \cite{Mc} using topological $\Z_2$ Deligne 
cohomology, a refinement of usual sheaf cohomology. 
The formula involves only the transition functions of the bundle 
relative to some trivializing open cover $\{U_i\}$ of $M$. For each point 
$x$ in $U_{i_0}\cap\cdots\cap U_{i_k}$, 
there is a $(k-1)$-cycle on 
${\rm SO}(n)/{\rm SO}(k-1)$ which depends continuously on $x$. 
The vertices of this cycle are defined by the value
at $x$ of the transition functions of the bundle.
A class in $H^k(M;\Z_2)$ is defined by 
sending $x$ to 1 if this cycle is homologically nontrivial, and to 0 
otherwise. 
This class coincides with 
$w_k(FE)$, where $FE$ is the frame bundle of $E$. 
Another description views the Stiefel-Whitney classes as representing 
obstructions to orientations with respect to generalized cohomology 
theories. The cases most relevant to this article, namely $w_4$ and 
$W_7$, are worked out in \cite{KS1}.

\vspace{3mm}
Now we recall some tools (see e.g. \cite{MT}).

\paragraph{Reduction in cohomology.}
Let $\rho_i: \Z \to \Z_j$ denote mod $j$ reduction, that is 
$\rho_j(1)=1$ mod $j$, for $j=2,3,\cdots$. We are mainly
interested in the case $j=2$, corresponding to mod 2
reduction, and the case $j=4$, corresponding to mod 4
reduction (see section \ref{Sec cong}). We use the same notation for the cohomology 
homomorphism induced by $\rho_i$. 
Let $\delta_*$ be the Bockstein coboundary associated with the 
exact sequence 
$
0 \to \Z \buildrel{k}\over{\longrightarrow} \Z
\buildrel{\rho_k}\over{\longrightarrow} \Z_k \to 0
$
(of coefficients) 
and set $\beta_k=\rho_k \delta_*$.

\paragraph{Mod 2 reduction.}
The coboundary $\beta_2$ is a 
derivation, that is for any two mod 2 cohomology classes $u$ and $v$, 
the relation
$
\beta_2(uv)= \beta_2(u) v + u \beta_2(v)$ holds.
The action of $\beta_2$ on the Stiefel-Whitney classes is 
\(
\beta_2(w_{2i})= w_1 w_{2i} + w_{2i+1}\;,
\qquad
\beta_2(w_{2i+1})= w_1 w_{2i+1}\;.
\)
We are dealing with oriented bundles, so the second equation will not
be of relevance to us, while the first equation reduces to 
$
\beta_2(w_{2i})= w_{2i+1}$.
An example of this, for $i=3$, is equation \eqref{w7} since $Sq^2 w_4=w_6$.

\paragraph{Action of Steenrod squares on Stiefel-Whitney classes.}
There are two flavors of the Steenrod square cohomology operation.
The first one is a mod 2 to integral cohomology operation
\(
Sq^i_\Z: H^k(X;\Z_2) \to H^{k+i}(X;\Z)\;,
\)
and the second one is a mod 2 to mod 2 cohomology operation
\(
Sq^i: H^k(X;\Z_2) \to H^{k+i}(X;\Z_2)\;.
\)
The two are related by $Sq^i_\Z= \beta Sq^{i-1}$.
The 
action of the Steenrod square on the Stiefel-Whitney classes of 
a bundle $E$ is given by the 
Wu formula 
\(
Sq^i w_{i+1}(E)= \sum_{j=0}^i w_j(E) w_{2i+1-j}(E)\;.
\label{wu}
\)
For $i=1,2,$ and 3, respectively, this gives
\begin{eqnarray}
Sq^1 w_2 &=& w_1 w_2 + w_3\;,
\nonumber\\
Sq^2 w_3 &=& w_5 + w_1 w_4 + w_2 w_3\;,
\nonumber\\
Sq^3 w_4 &=& w_7 + w_1 w_6 + w_2 w_5 + w_3 w_4\;.
\label{eq wu}
\end{eqnarray}
Note that when $E$ is oriented and Spin, we have
$w_3(E)=0$, $w_5(E)=0$ and so $Sq^3w_4(E)=w_7(E)$. 
It is generic that the interesting even degree 
classes are the mod 2 ones,
$w_{2k}$, while interesting odd degree
classes are images of these in integral 
cohomology, i.e. $Sq_\Z^iw_{2k}=W_{2k+i}$ (for appropriate $i$).

\paragraph{Spin characteristic classes}
When a bundle $E$ is Spin the corresponding characteristic 
classes will admit special values. The second Stiefel-Whitney class
$w_2(E)$ will vanish, and the first Pontrjagin class $p_1(E)$ will 
be divisible by two because of the relation $p_1(E)=w_2(E)^2$ mod 2. 
Then, naturally, the bundle would be 
described via characteristic classes related to the Spin group
rather than to
the orthogonal group. 
The integral cohomology ring of the classifying space of the 
Spin group is \cite{T}
\(
H^*(B{\rm Spin};\Z) \cong \Z[Q_1, Q_2, \cdots] \oplus T\;,
\label{tho}
\)
where the generators
$Q_i\in H^{4i}(B{\rm Spin};\Z)$ 
are the Spin characteristic classes, defined via the 
corresponding Pontrjagin classes by
\(
p_1= 2 Q_1\;, \qquad
p_2= Q_1^2 + 2Q_2\;.
\label{Q1}
\)
The summand $T$ in \eqref{tho}
is 2-torsion, $2T=0$. For example, in dimension 
seven this is generated by the seventh Stiefel-Whintey class
$W_7=\delta w_6$, with $\delta: H^*(X;\Z_2) \to H^*(X;\Z)$ 
the Bockstein on cohomology. 
Mod 2 characteristic classes of 
Spin bundles are obtained by pullback from the universal 
classes generating the cohomology ring of BSpin \cite{Qu}
\(
H^*(B{\rm Spin};\Z_2)\cong \Z_2[ w_4, w_6, w_7, w_8, w_{10},\cdots]\;.
\)
Obviously, all classes $w_i$ up to $i=3$ are absent. 
The degree four Spin characteristic class $Q_1=\lambda:=\frac{1}{2}p_1$ 
admits $w_4(E)$ as a mod 2 reduction 
\(
\rho_2(Q_1(E)) = w_4(E)\;.
\label{r2w4}
\)
Generally, the classes $Q_i$ admit the Stiefel-Whitney classes
in the same dimension as mod 2 reductions
$
\rho_2 (Q_i)= w_{4i}$.
Here $\rho_2$ is the induced homomorphism 
on cohomology arising from the corresponding 
reduction in coefficients $\rho_2: \Z \to \Z_2$. 
We will see more of this in section \ref{Sec cong}.

\section{(Twisted) Membrane Structures}
\label{t M2}

We now define the first new structure considered in this paper. 

\paragraph{Membrane structure.}
The Stiefel-Whitney class 
$w_4$ is the mod 2 reduction of $\lambda=\frac{1}{2}p_1$. This implies that 
$\lambda$ is even
if and only if $w_4=0$. 
%
%
We use this as the obstruction to having a {\it Membrane structure}; that is 
a Membrane structure on a  bundle $E$  can be defined when $w_4(E)=0 \in H^4(E;\Z_2)$. This is closely related to a String structure in the following sense. If 
$\lambda (E)=0$ then $w_4(E)=0$; this means that String bundles are 
automatically Membrane bundles.  However, certainly there are bundles $E$
for which $w_4(E)=0$ but $\lambda (E)\neq 0$; for instance, $\lambda (E)$ 
 instead is an even 
class.

\paragraph{Additive structure in the untwisted case.}
If at least one of two real bundles $E$, $F$, is oriented and Spin,
then for the Whitney sum we have
\begin{eqnarray}
w_4(E \oplus F)&=&
w_4(E) + w_3(E) w_1(F) + w_2(E) w_2(F) + w_1(E) w_3(F) + w_4(F)
\nonumber\\
&=& w_4(E) + w_4(F)\;.
\end{eqnarray}
Here we use the fact that the first nontrivial Stiefel-Whitney class 
should be in even degree (cf. \eqref{wu}).
Therefore, in particular, the product of 
two oriented Spin Membrane manifolds is again an 
oriented Spin Membrane manifold.

\paragraph{Application: Membrane structures related to the M5-brane.}
Consider the M5-brane with worldvolume $\cW^6$ embedded
in eleven-dimensional spacetime $Y$, with normal bundle 
$\mathcal{N}$. Let $S(\mathcal{N})$ be the unit sphere 
bundle of $\mathcal{N}$
of dimension ten 
and $\pi: S(\mathcal{N}) \to \cW^6$ the projection. Let $a$ be a degree 
four class on $S(\mathcal{N})$. Then \cite{W-Duality}
\(
\pi_*(a \cup a) \cong w_4(\mathcal{N})\quad {\rm mod~}2\;.
\label{a cup a}
\)
It is desirable that the left hand side of \eqref{a cup a} be
even so that the partition function is well-defined \cite{W-Duality}.
We see that this condition is satisfied when $w_4(\mathcal{N})=0$,
i.e. if the normal bundle to the M5-brane admits a Membrane 
structure. One can also get such a structure on the worldvolume,
under some conditions. For example, if $\cW^6$ is a
 2-connected six-manifold then 
we have $H^4(\cW^6;\Z_2)=0$. In particular, $w_4(\cW^6)=0$, 
so that indeed the M5-brane 
worldvolume admits a Membrane structure.

\vspace{3mm}
Next, we consider the twist for the membrane structure, in analogy to other 
structures \cite{Wa} \cite{SSS3}.

\paragraph{The definition.}
Let $(X, \alpha)$ be a compact topological space with a 
degree four cocycle $\alpha: X \to K(\Z_2, 4)$. An 
{\it $\alpha$-twisted Membrane-manifold} over $X$ 
is a quadruple $(M, \nu, \iota, \eta)$, where

\noindent (1) $M$ is a smooth compact oriented manifold together with a 
fixed classifying map of its stable normal bundle
$
\nu : M \to B{\rm SO}$;

\noindent (2) $\iota: M \to X$ is a continuous map;

\noindent (3)  $\eta$ is an $\alpha$-twisted Membrane-structure on 
$M$, that is, a homotopy commutative diagram

  \(
    \raisebox{20pt}{
    \xymatrix{
       M
       \ar[rr]^\nu_>{\ }="s"
       \ar[d]_\iota
       &&
       B{\rm SO}
       \ar[d]^{w_4}
       \\
       X
       \ar[rr]_\alpha^<{\ }="t"
       &&
       K(\Z_2,4)
       \ar@{=>}^\eta "s"; "t"
    }
    }
    \label{a t M2}
    \,,
\)
where $w_4$ is the classifying map of principal 
$K(\Z_2, 3)$ bundles associated to the fourth 
Stiefel-Whitney class, and $\eta$ is a homotopy 
between $w_4 \circ \nu$ and $\alpha \circ \iota$. 


\paragraph{Remarks.} Given a smooth compact oriented
$n$-manifold $M$ and a topological space $X$ with a 
twisting $\alpha: X \to K(\Z_2, 4)$, then
\begin{enumerate}
\item $M$ admits an $\alpha$-twisted Membrane-structure if and only if
there exists a continuous map $\iota: M \to X$ such that
\(
w_4(M)+\iota^*([\alpha])=0\in H^4 (M;\Z_2)\;. 
\label{w4 a}
\)
\item If \eqref{w4 a} is satisfied, then the set of equivalence classes of 
$\alpha$-twisted Membrane-structures on $M$ are in one-to-one 
correspondence with elements in $H^3(M;\Z_2)$.  

\item If the twisting $\alpha: X \to K(\Z_2, 4)$ is 
homotopic to the trivial map then an $\alpha$-twisted 
Membrane structure on $M$ is equivalent to a 
Membrane structure on $M$. 

\item Two $\alpha$-twisted Membrane structures 
$\eta$ and $\eta'$ on $M$ are equivalent if there is 
a homotopy between $\eta$ and $\eta'$. 

\end{enumerate}

\paragraph{Additive structure in the twisted case.}
Let $\alpha$ be the same as in diagram \eqref{a t M2}.
 An example of such an $\alpha$ occurs by considering 
fractional $G_4$ flux.
The same result as in the untwisted case holds. That is,
for a product manifold $X$ whose tangent bundle splits as $TX=E \oplus F$ and with a
twisting $\alpha: X \to K(\Z_2, 4)$,
we have
\begin{eqnarray}
(w_4+ \alpha)(E \oplus F)&=& 
w_4(E \oplus F) + \alpha(E \oplus F)
\nonumber\\
&=& w_4(E) + w_4(F) + \alpha(E) + \alpha(F)
\nonumber\\
&=&
(w_4 + \alpha)(E) + 
(w_4 + \alpha)(F)\;.
\label{eq wa}
\end{eqnarray}
 Therefore, the product of two oriented Spin twisted Membrane 
manifolds is also an oriented Spin twisted Membrane 
manifold.

\paragraph{Examples.} 
A manifold $M$ is a boundary if and only if all its Stiefel-Whitney 
numbers vanish (see \cite{Stong}). Since every flat manifold is a boundary
\cite{HR}, 
all Stiefel-Whitney numbers of a flat manifold vanish. 
However, the same is not true for the corresponding 
classes. Indeed, there are examples of flat manifolds 
with non-vanishing Stiefel-Whitney classes. 
Consider toral extensions, i.e. torus bundles over
flat manifolds, which can certainly arise in 
realistic physical models (cf. \cite{DMW-Spinc}),
have all the even-dimensional Stiefel-Whitney classes
nonzero up to the middle dimension. So in the case
of string theory or M-theory, we get nonvanishing 
of $w_2$ and $w_4$. Such manifolds can be desribed 
as follows. Let $Q\subset \R^n \ltimes O(n)$ be a Bieberbach group so that 
it acts on $\R^n$ freely and properly discontinuously. Suppose $Q$ also 
acts on the torus $T^k$ as isometries so that it acts on the product 
$T^k \times \R^n$ diagonally as isometries. The quotient $M=(T^k \times 
\R^n)/Q$ is called \cite{V} a {\it flat toral extension} of the compact 
flat Riemannian manifold $N=\R^n/Q$. 
Then $M$ is a torus bundle over $N$.

\vspace{3mm}
There are other examples which are not toral extensions \cite{IK}. There 
is a class of $(2n+1)$-dimensional {\it compact} flat manifolds whose 
Stiefel-Whitney classes $w_{2j}$ are non-zero for $0\leq 2j \leq n$, none 
of which is a flat toral extension of another flat manifold. 
This manifold  $M$ has holonomy group 
$(\ZZ_2)^{n+1}$ and has a vanishing first Betti number $b_1(M)=0$.
Taking $n=5$ we get eleven-manifolds with nonzero 
$w_2$ and $w_4$ with the above properties. 

\paragraph{Interpretation of the $w_4$ condition.}
The $w_4$ condition arises as an orientation condition with 
respect to $EO(2)$-theory, needed to construct an anomaly-free
partition function in type IIA string theory \cite{KS1}. 
On the other hand,
Stiefel-Whitney class of dimension close to the dimension of the 
space take on interesting roles. For example, 
when the manifold $M$ is 4-dimensional, $w_4$ admits 
a special interpretation, related to 
global causality. In \cite{FA}
it is shown that if spacetime $(M,g)$ is stably causal, then 
$w_4(M)=0$, and there exists a 5-manifold $V$ such that $M=\partial V$.

\paragraph{Application: Twisted Membrane structure associated to the M2-brane.}
If $\lambda (M)$ is not even then $w_4(M)\neq 0 \in H^4(M;\Z_2)$. 
Then there exists a class $\alpha$ (related to $G_4$, cf. expression \eqref{flux}) such that
$w_4 + \alpha=0\in H^4(M;\Z_2)$.   
This is the case for the quantization of the C-field when 
$\lambda$ is not divisible by two. 
Therefore, a twisted Membrane structure arises naturally from considering M-theory on manifolds with an odd first Spin class. 

%
%
%
%
%
%

\paragraph{Application: Twisted Membrane structure associated to the M5-brane.}
Consider the situation of M-theory on 
$\R^5/\Z_2$, giving rise to an M5-brane \cite{Flux}. 
A four-cycle surrounding the 
origin in this space can be taken to be 
the real projective space $\R P^4=S^4/\Z_2$. 
The mod 2 cohomology ring of $\R P^4$ is 
a polynomial ring in a degree one generator
$x$ with relation $x^5=0$. This gives that 
$x^4$ is the mod 2 fundamental class of
$\R P^4$, so that $\int_{\R P^4} w_4=1$ mod 2.
Thus, there is a half-integral flux of 
$G_4$. 
A similar situation arises when considering 
the $\R^5/\Z_2$ orbifold with a $\Z_2$ fixed plane 
along a Riemann surface 
$\Sigma$ \cite{Hor}. This gives the integral over the 
relevant 4-cycle $\int_{S} w_4= \chi (L)$ mod 2,
where $\chi (L)$ is the Euler characteristic 
of a line bundle over $\Sigma$. We can see that 
when $\chi (L)$ is odd then $w_4=1$. 
In both cases, we get that there exists a class 
$\alpha$ such that $w_4 + \alpha=0 \in H^4(M;\Z_2)$,
that is we have a twisted Membrane structure.

\paragraph{Boundary case.}
Let $(M, \nu, \iota, \eta)$ be an $\alpha$-twisted Membrane 
manifold over $X$. Then there is a natural $\alpha$-twisted
Membrane structure on the boundary $\partial M$ with outer
normal orientation which is the restriction of the 
$\alpha$-twisted Membrane structure on $M$

 \(
    \raisebox{20pt}{
    \xymatrix{
    \partial   M
       \ar[rr]^{\nu|_{\partial M}}_>{\ }="s"
       \ar[d]_{\iota|_{\partial M}}
       &&
       B \mathrm{Spin}
       \ar[d]^{w_4}
       \\
       X
       \ar[rr]_\alpha^<{\ }="t"
       &&
       K(\Z_2,4)
       \ar@{=>}^{\eta_{\partial M}} "s"; "t"
    }
    }
    \,.
    \label{bdry}
\)

\paragraph{Examples.} 
M-theory can be formulated on a manifold with boundary,
and the quantization condition for the $C$-field 
extends to the boundary (cf. \cite{DFM}). 
Hence, we can consider the structures we define in this paper 
restricted to that boundary. Similarly for the M2-brane and 
the M5-brane; they admit boundaries on the M5-brane and 
the M9-brane, respectively (see \cite{St} \cite{To}).
In addition to the case of the twisted Membrane structure 
explicitly considered above,  
the other structures restrict to the boundary 
in a similar way,  with the obvious changes to 
diagram \eqref{bdry}.

\paragraph{Remarks on orientation.}
{\bf 1.} Let $\tau_X: X \to B{\rm Spin}$ be the classifying map 
of the stable tangent bundle of $X$. Then 
a $w_4 \circ \tau_X$-twisted Membrane structure on $M$ is 
equivalent to an $EO(2)$-oriented map 
from $M$ to $X$.

\noindent {\bf 2.} Let $(M, \nu, \iota, \eta)$ be an $\alpha$-twisted 
Membrane manifold over $X$. Any 
$EO(2)$-oriented map $f: M' \to M$ 
defines a canonical $\alpha$-twisted 
Membrane structure on $M'$.

\section{(Twisted) Membrane${}^c$ Structures}
\label{t M2 c}

\paragraph{Membrane${}^c$ structures.}
We define a Membrane${}^c$ structure in an analogous way to a 
Spin${}^c$ structure, where instead of using $W_3$ we use $W_5=\beta w_4$. 
Recall that a Membrane structure is defined when the obstruction 
$w_4$ vanishes. Now if this obstruction is not zero but is a mod 2 reduction 
of an integral class (which would be $Q_1$ in the Spin case) then 
$W_5$  is zero. We take this as defining the obstruction to
having a Membrane${}^c$ structure. 


\vspace{3mm}
We now consider the twisted case. 

\paragraph{Twisted Membrane${}^c$ structures.} 
Let $f: X \to B{\rm O}$ be the classifying map for the orthogonal bundle over $X$.
An $H_5$-twisted 
Membrane${}^c$ structure on a space $X$ is defined by the 
homotopy commutative diagram
\(
    \xymatrix{
       X
       \ar[rr]^f_>>>{\ }="s"
       \ar[drr]_{H_5}^{\ }="t"
       &&
       B \mathrm{O}
       \ar[d]^{W_5}
       \\
       &&
       K(\mathbb{Z},5)
       \ar@{=>}^\eta "s"; "t"
    }
    \,,
  \)
where the homotopy $\eta$  is between the map $f \circ W_5$ ($f^* W_5=W_5(X)$) 
and the 
five-cocycle corresponding to the class $H_5\in H^5(X;\Z)$. 
We can also include a brane as in diagram \eqref{a t M2}
and the extension is obvious.

\paragraph{Example/Application.}
We consider a Membrane${}^c$ structure with a trivial twist. 
The class $W_5(Y) \in H^5(Y;\Z)$ is obtained from the Bockstein 
homomorphism applied to $w_4(Y)$. The degree five class is interpreted in 
\cite{DFM} as the background magnetic charge induced by the 
topology of $Y$, and must vanish 
in order to be able to formulate any (electric) $C$-field. 
When $Y$ is Spin, $W_5(Y)=0$, since the class $\lambda$ is 
an integral lift of $w_4(Y)$.  
Similarly, the obstruction to existence of a $Sp(2)$ 
bundle and to a global Spin$(1,5)$ bundle 
on the M5-brane wolrdvolume $\cW^6$
cancel if $W_5(\cW^6)=0$ (cf. \cite{Ka}).

\section{Twisted String${}^{K(\Z,3)}$ Structures}
\label{String KZ3}

\paragraph{String${}^{K(\Z,3)}$ structures.}
These structures are defined 
using the seventh Stiefel-Whitney class. Let us start by 
mentioning a distinction.
For the mod 2 Stiefel-Whitney classes we have
\(
Sq^1 Sq^2 w_4= Sq^3 w_4= w_7 \in H^7 \Z_2\;,
\label{w7}
\)
while
for the Spin characteristic class $\lambda=Q_1$ we have
\(
Sq^3 \lambda = Sq^3 Q_1= W_7 \in H^7\Z\;.
\)
We define a String${}^{K(\Z,3)}$ structure on a manifold 
$M$ by the condition
$W_7(M)=0$. 

\paragraph{Application.} The DMW anomaly \cite{DMW} 
for dimensionally-reduced M-theory partition function to be 
well-defined 
is given exactly by $W_7=0$. This is discussed extensively 
in \cite{KS1}.

\paragraph{Examples.}
 {\bf 1.} An example which is a String${}^{K(\Z, 3)}$-manifold but
 not a String manifold is $X^{10}=S^2 \times S^2 \times \C P^3$.
 This has a nonzero $\lambda$ (non-torsion), while there
 is no odd cohomology, so that $W_7=0$.

\noindent {\bf 2.} An example which is neither String nor String${}^{K(\Z,3)}$ is
the eight-dimensional Spin homogeneous space of Lie groups
$G_2/SO(4)$. 
This has nonzero Stiefel-Whitney classes $w_4$, $w_6$ and 
$w_8$ (see \cite{BH1}). The class $w_6$
gives that $W_7(G_2/SO(4))\neq 0$, so 
that the space is not String${}^{K(\Z,3)}$. 
On the other hand, the Pontrjagin classes are
$p_1(G_2/SO(4))=2$ and $p_2(G_2/SO(4))=7$, so that 
the first Spin characteristic class is $Q_1(G_2/SO(4))=1$
(indeed $w_4(G_2/SO(4))=1$). 
This implies that $G_2/SO(4)$ is not String.

\paragraph{Additive structure in the untwisted case.}
We consider the class of a Whitney sum of two bundles $E$ 
and $F$,
\begin{eqnarray}
W_7(E \oplus F)&=&
\beta w_6 (E \oplus F)
\nonumber\\
&=&
\beta \left[ 
w_6(E) + w_5(E) w_1(F) + w_4(E) w_2(F) +
w_3(E) w_3(F)+ \right.
\nonumber\\
&&\left. ~~~~~~~~~~~~~~~~~~~~~~~+ w_2(E) w_4(E) + w_1(E) w_5(F) + w_6(F)
\right]\;.
\label{w exp}
\end{eqnarray}
\begin{enumerate}
\item If $E$ and $F$ are both oriented then $w_1(E)=0=w_1(F)$ and 
 the two terms containing $w_1$ will vanish.

\item If $E$ and $F$ are both Spin then, in addition, $w_2(E)=0=w_2(F)$, 
so that all cross-terms in \eqref{w exp} vanish, leaving
\(
W_7(E \oplus F)= W_7(E) + W_7(F)\;.
\label{W7 ad}
\)
Here we used the 
 the Wu formula \eqref{eq wu}.
\item The same conclusion holds if $E$ and $F$ are not Spin but only 
$EO(2)$-orientable, that is if $w_4(E)=0=w_4(F)$. 
\end{enumerate}

\paragraph{Example.}
Consider a product manifold $Z=X\times Y$. Then the tangent bundle of
$Z$ is the Whitney sum of the tangent bundles of $X$ and $Y$, so that
we can apply \eqref{W7 ad}. If neither $X$ nor $Y$ have a String${}^{K(\Z,3)}$
structure, then this means that $W_7(X)= W_7(Y)=1$, and \eqref{W7 ad} gives
that $W_7(Z)=0$, since $W_7$ is 2-torsion. This then says that out of two 
manifolds neither of which is String${}^{K(\Z,3)}$, we can build a third manifold
in a straightforward way, namely their product, which is String${}^{K(\Z,3)}$.

\vspace{3mm}
\noindent Next we consider a twist for a String${}^{K(\Z,3)}$ structure.

\paragraph{The definition.}
Let $(X, H_7)$ be a compact topological space with a 
degree seven cocycle $H_7: X \to K(\Z, 7)$. An
{\it $H_7$-twisted String${}^{K(\Z, 3)}$-manifold} over $X$ 
is a quadruple $(M, \nu, \iota, \eta)$, where

\noindent (1) $M$ is a smooth compact Spin manifold together with a 
fixed classifying map of its stable normal bundle
$
\nu : M \to B{\rm Spin}$;

\noindent (2) $\iota: M \to X$ is a continuous map;

\noindent (3) $\eta$ is a $H_7$-twisted String${}^{K(\Z,3)}$-structure on 
$M$, that is, a homotopy commutative diagram

  \(
    \raisebox{20pt}{
    \xymatrix{
       M
       \ar[rr]^\nu_>{\ }="s"
       \ar[d]_\iota
       &&
       B \mathrm{Spin}
       \ar[d]^{W_7}
       \\
       X
       \ar[rr]_{H_7}^<{\ }="t"
       &&
       K(\mathbb{Z},7)
       \ar@{=>}^\eta "s"; "t"
    }
    }
    \,.
\)


\paragraph{Additive structure in the twisted case.}
Now consider the twisted class $W_7^H:=W_7 + H_7$. Consider the 
value on the Whitney sum of two bundles $E$ and $F$. 
Similarly to the case of twisted Membrane structures (cf. equation \eqref{eq wa}),
for a product manifold $X$ with a splitting of the tangent bundle $TX=E \oplus F$,
this gives
\(
W_7^H(E \oplus F) = W_7^H(E) + W_7^H(F)
\)
if the conditions as in the twisted case above are satisfied.

\paragraph{Application.}
The $H_7$ twist appears at the level of de Rham cohomology
in the dual formulation of 
heterotic string theory \cite{7}. 
Since elliptic genera are needed to calculate anomalies in 
heterotic string theory (see \cite{LNSW}), 
and since dualities connect this 
theory to type II theories and M-theory where elliptic refinements 
occur \cite{KS1}\cite{KS2} \cite{KS3}, this all strongly 
gives that elliptic cohomology is a natural tool to study the heterotic 
theory. A further indication is the 
requirement of a string structure \cite{W-White} \cite{Kil}. 
Then, the existence of the above twist should 
admit a lift to generalized cohomology \cite{7}, where
the condition $W_7 + H_7$ would appear. 
Note also that we have the twisted String${}^{K(\Z,3)}$
structure already from the twisted String condition
(albeit in a trivial way), as we show in section 
\ref{relate}.

\section{Twisted Fivebrane${}^{K(\Z,4)}$ Structures}
\label{Fivebrane KZ4}

\paragraph{Fivebrane${}^{K(\Z,4)}$ structures.}
Recall that in \cite{II} we interpreted a String${}^c$ structure 
as a twisted String structure with a product twist, i.e. 
that is coming from the cup product of two elements from 
$K(\Z, 2)$. So then a String${}^c$ structure might 
alternatively be called a String${}^{K(\Z,2)}$ structure.
Note that the twist is still a degree four class, but
taken to be a composite, i.e. built out of a product of 
two copies of (the same) degree two element. 

\vspace{3mm} 
Similarly in the Fivebrane case, we can define 
a Fivebrane${}^{K(\Z,4)}$ structure to correspond to 
a twisted Fivebrane structure where the degree eight
twist is a composite of two degree four twists. 

\vspace{3mm}
We need the following for the definition below  (see \cite{Mau} or \cite{May}).
For any based space $X$, the suspension with the $S^0=\{0,1\}$ 
gives $S^0 \wedge X=X$. This shows that every spectrum is canonically 
a module over the sphere spectrum. Taking $X$ to be the Eilenberg-MacLane
spectrum $(H\Z)_n=K(\Z,n)$, gives that $S^0 \wedge K(\Z, n)=K(\Z,n)$.
This gives rise to a morphism $s: S^0 \to K(\Z,n)$ which,
together with the identity map ${\rm id}: K(\Z, n) \to K(\Z,n)$, 
gives a product map $s \times {\rm id}: S^0 \times K(\Z, n) \to 
K(\Z, n) \times K(\Z,n)$. This is turn induces a map 
$S^0 \wedge K(\Z, n) \to K(\Z, n) \wedge K(\Z, n)$, which 
in turn gives a map $\wedge: K(\Z, n) \to K(\Z,n) \wedge K(\Z,n)$.

\paragraph{The definition.}
A {\it Fivebrane${}^{K(\Z,4)}$ structure} on a space $X$ with a
String structure classifying map $f$ with a degree eight
cocycle $\alpha$ is characterized 
by homotopy between the String class $\frac{1}{6}p_2$
and the composite cocycle $\alpha$. 
 The cocycle $\alpha$  is a cup product of two degree four cocycles
 defined via the
 map 
$
l_4: K(\Z, 4) \times K(\Z,4) \to K(\Z, 8)
$,
which classifies the cup product operation 
$
H^4(X;\Z) \times H^4(X;\Z) \to H^8(X;\Z)$.
More precisely, we have 
the diagram 
\(
\xymatrix{
X
\ar[rr]^f_>{\ }="s"
\ar[d]_{a(X)}
\ar[ddrr]^\alpha^<{\ }="t"
&&
B{\rm String}(n) 
\ar[dd]^{\frac{1}{6}p_2}
\\
K(\Z,4) 
\ar[d]^{\wedge}="a"
&&
\\
{K(\Z,4) \wedge K(\Z,4)}
\ar[rr]^{\cup}="b"
&&
K(\Z,8)
\ar@{=>}^{\eta_1} "s"+(2,-2); "s"- (8,8)
\ar@{<=}_{\eta_2} "a"+(2,-2) ; "a" +(12,4)
}\;.
\label{C D}
\)
The first homotopy $\eta_1$ gives the relation
$\frac{1}{2}p_2 + \alpha=0 \in H^8(X;\Z)$ and the second 
homotopy $\eta_2$ gives $\alpha + \frac{1}{2}a^2=0 \in H^8(X;\Z)$.
Combined, the two homotopies then give
\(
\frac{1}{6}p_2 + \frac{1}{2}a^2=0 \in H^8(X;\Z)\;.
\label{obs 5kz4}
\)
This identifies a Fivebrane${}^{K(\Z,4)}$ 
structure as a special case of 
a twisted Fivebrane structure of \cite{SSS3}.

\paragraph{Application: The dual $C$-field in M-theory.}
The equation of motion for the $C$-field in M-theory
has an electric source which is an eight-form. In the 
case when the background manifold admits a Spin 
structure -- which is the simplification assumed 
also in \cite{SSS3}-- the dual class is (cf. \eqref{dual flux}) 
\(
\frac{1}{48}p_2(Y) + \frac{1}{2} a^2\;.
\label{dualc}
\)
Note the extra factor of 8 in the denominator of the coefficient of
$\frac{1}{6}p_2$ in \eqref{dualc}
in comparison to the obstruction to the Fivebrane${}^{K(\Z,4)}$
structure \eqref{obs 5kz4}. The structure described 
by \eqref{dualc} should be an 
$\mathcal{F}_{\langle 8 \rangle}^{K(\Z,4)}$ structure. 
Here $\mathcal{F}_{\langle 8 \rangle}$ is the object fibering 
over BString whose class is $\frac{1}{48}p_2$ instead of 
$\frac{1}{6}p_2$, as explained in \cite{SSS2}\cite{SSS3}. 
An alternative point of view would be to perform 
reductions mod 8 via cohomology operations. We will 
do the analog for the String case next in section \ref{Sec cong}.

\section{Congruences for String and Fivebrane classes}
\label{Sec cong}

The quantization of the C-field and the Green-Schwarz anomaly
cancellation involves a fractional first Pontrjagin class, and similarly
the dual C-field and the dual Green-Schwarz cancellation involves
a fractional second Pontrjagin class \cite{SSS2}\cite{SSS3}.
Furthermore, these classes do not precisely match 
String and Fivebrane structures, respectively, but only up to 
further divisions \cite{SSS2} \cite{SSS3}.
Here we study such division (or fractions) from the point of view of 
cohomology operations. 

\subsection{String class mod 2}
\label{String 2}

In this section we will consider the class 
$\frac{1}{2}\lambda=\frac{1}{2}Q_1=\frac{1}{4}p_1$ appearing the flux 
quantization condition of the C-field \eqref{flux}.
For this, we will study the mod 4 reduction of 
the first Pontrjagin class. 
We start with setting up some notation and providing some
basic definitions.

\paragraph{Mod 4 reduction.}
The epimorphism $\rho_4: \Z \to \Z_4$ induces a homomorphism 
of cohomology groups 
$
H^*(X;\Z) \longrightarrow  H^*(X;\Z_4)$, 
$\alpha \longmapsto  \rho_4(\alpha)$.
Let us illustrate this for the Pontrjagin classes.
A theorem of Wu says that for any $O(m)$ bundle $E$ over 
$X$, the class $\rho_4(p_k(E)) \in H^{4k}(X;\Z_4)$ is determined 
by the Stiefel-Whitney classes $w_l (E) \in H^l(X;\Z_2)$. In particular,
if the Stiefel-Whitney classes $w_1(E), \cdots , w_{k-1}(E)$ are
zero then $\rho_4(p_k(E))= i_{2*}w_{4k}(E)$, 
where $i_2$ is the homomorphism induced by the inclusion of 
coefficients
$i_2 : \Z_2 \to \Z_4$.

\paragraph{Pontrjagin squares.}
Next, let the inclusion $i_j: \Z_2 \to \Z_{2^j}$, $i \in \mathbb{N}$,
 be defined  by $i_j(1 ~{\rm mod~}2)= 2^{j-1}$ mod $2^j$. 
We are interested in mod $2^m$ reduction of cohomology classes.
The relevant cohomology operation is the 
Pontrjagin square. This is the cohomology operation (cf. \cite{MT})
\(
\mathfrak{P}: H^{2k}(X;\Z_2) \to H^{4k}(X;\Z_4)\;,
\)
satisfying the relation
$
\mathfrak{P} \rho_2(x) = \rho_4 (x^2)
$,
 for $x \in H^{2k}(X;\Z)$.

\vspace{3mm}
We will explore the mod 4 reduction of the first Pontrjagin class in various 
situations. M-theory can be considered on Spin${}^c$ manifolds, and the 
flux quantization extends. We can also consider complex manifolds
(as factors) as well as complex vector bundles. Note that the quantization 
condition extends also to the worldvolume of the M5-brane, which need not
be Spin or even Spin${}^c$ (see \cite{W-Duality} \cite{II} for discussions on these matters).

\paragraph{Mod 4 reduction for characteristic classes of bundles.} Let $E$ be a bundle
with some characteristic class(es).

\noindent {\it 1. Real bundles.}
First take $E$ to be an $n$-dimensional real vector bundle over a space $X$. 
Then the mod 2 reduction of its Pontrjagin class is 
\(
\rho_2(p_1(E))= w_2(E)^2\;.
\) 
On the other hand, the mod 4 reduction satisfies 
$
\mathfrak{P}(w_2)= \rho_4(p_1) + i_2 [ w_1 Sq^1 w_2 +  w_4]
$,
where $i_2: \Z_2 \to \Z_4$ is given by $i_2(1 {\rm ~mod~} 2)=2$ mod 4.
If the bundle $E$ is oriented, then 
\(
\mathfrak{P}(w_2)= \rho_4(p_1) + i_2 (w_4)\;.
\label{main w2}
\)
Note that if $w_2=0$ then $\rho_4(p_1)$ is given by 
$i_2(w_4)$. However, in the resulting Spin case
it is more appropriate to use Spin characteristic classes, 
as we do shortly -- see \eqref{r2w4}.


\noindent {\it 2. Complex bundles.}
If $E$ is a complex bundle with underlying real bundle $E_\R$ and 
second Chern class $c_2(E)$, then the mod 2 reduction gives
$
\rho_2(c_2(E))= w_4(E_\R)$.
The mod 4 reduction is simpler than in the real case, namely
\(
\mathfrak{P} (w_4)= \mathfrak{P}(\rho_2(c_2))= \rho_4(c_2^2)\;.
\)
\noindent {\it 3. Real forms.}
Now consider a bundle $E$ with a real form $E_\R$ with corresponding 
classes $c=c(E)$, $p=p(E_\R)$, and $w=w(E_\R)$. Since
$p_1(E_\R)= 2c_2(E) + c_1(E)^2$, then 
$
\rho_4(c_1^2)= \rho_4(p_1) - \rho_4(2 c_2)$. 

\paragraph{Relating mod 4 to mod 2 reduction.}
The mod 4 reduction can be related to the mod 2 reduction 
through the induced factor map $\kappa_2: \Z_4 \to \Z_2$ 
in the exact sequence $0 \to \Z_2 \buildrel{i_2}\over{\longrightarrow}
\Z_4 \buildrel{\kappa_2}\over{\longrightarrow} \Z_2 \to 0$ as
$\kappa_2 \rho_4=\rho_2$. 
Then 
$
\pm 2 \rho_4(c_2)= i_2 \rho_2(c_2)= i_2(\rho_2(c_2))= i_2(w_4)\;,
$
so that the mod 4 reduction of $p_1$ satisfies the relation
\(
\rho_4(c_1^2)= \rho_4(p_1) + i_2(w_4)\;.
\)

\paragraph{Special values in dimensions 4 and 8.}
Now consider an $O(m)$ bundle $E$ over the sphere 
$S^{4k}$. Then the Pontrjagin class $p_k(E) \in H^{4k}(S^{4k};\Z)$ 
is divisible by $\epsilon(2k-1)!$, with $\epsilon$ being 1 in
dimensions $4k=8m$ and 2 in dimensions $4k=8m+4$. 
So, the first and second Pontrjagin 
classes of $E$ are divisible by 2 and by 6, respectively. 
These give rise to generators of BSpin and BString, 
respectively.
Since 
\(
\rho_4(p_k(E))= i_{2*} w_{4k}(E) \in H^{4k}(S^{4k};\Z_4) =\Z_4\;,
\)
the class $w_{4k}(E)$ is zero if and only if $p_k(E)$ is divisible by 4. 
This 
is related to parallelizability and to division algebras \cite{M58}.

\paragraph{Dimension less than eight.}
Expression \eqref{main w2} is a relation involving the 
Stiefel-Whitney classes $w_2$ and $w_4$. If we 
find another relation among these two classes then 
the original expression would then simply relate one
Pontrjagin class to one Stiefel-Whitney class. 
Such simplifications happens in relatively low dimensions.
In general, the 4th Wu class is related to the Stiefel-Whitney 
classes via
\(
v_4=w_4 + w_2^2\;.
\)
Note that the Wu class is relevant in relating M-theory to twisted K-theory in 
type IIA string theory \cite{BM}, where it is required to admit
a  lift to twisted cohomology. 
Now if we are in a situation where $v_4=0$ then this gives the 
desired relation between $w_2$ and $w_4$ (assuming 
neither is zero). This 
occurs, for example, for any oriented closed manifold of 
dimension less than eight (i.e. when the second 
Wu class does not appear for dimension reasons). Of course
one could also consider other situations in higher dimensions 
where this could still happen. Replacing 
$w_4$ by $w_2^2$ in \eqref{main w2} gives
\bea
\rho_4(p_1)&=& \mathfrak{P}(w_2) + i_2(w_2^2)
\nonumber\\
&=&
\mathfrak{P}(w_2) + 2 \mathfrak{P}(w_2)\;,
\eea
so that the new relation for the mod 4 reduction of $p_1$ is 
\(
\rho_4(p_1)= - \mathfrak{P} (w_2) \quad {\rm when ~} v_4=0.
\)


\paragraph{Spin${}^c$ bundles.} For Spin${}^c$ bundles, 
like real oriented bundles, the mod 4 reduction gives
\(
\rho_4(p_1(E))= \mathfrak{P} w_2(E) + i_2(w_4(E))\;.
\)

\subsection{Fivebrane class mod 8}
In this section we will consider the class $\frac{1}{48}p_2$
appearing in the quantization of the dual of the C-field 
\eqref{dualc}.
The discussion here is analogous to the case of 
the C-field in section \ref{String 2}; instead of reducing 
mod 2 we should reduce mod 8. However, we found that 
this case requires extensive discussion and so we will 
leave it to a separate treatment, and 
allow ourselves to be content here with the analogy with the 
String case and with having a novel way of writing the 
corresponding class. 

\paragraph{String characteristic classes.}
A bundle $E$ with a String structure is characterized by the vanishing 
of the first Spin characteristic class $Q_1(E)=0$ (and how it vanishes since it 
is a homotopy). This has implications
on the Stiefel-Whitney classes. Since $\rho_2(Q_1)=w_4$, this implies
that $w_4(E)=0$. Furthermore, since $Sq^3 Q_1=W_7$, this 
implies in addition that $W_7(E)=0$.  Analogously to the Spin case,
we naturally seek to characterize the corresponding String bundles 
with characteristic classes of  BString. Here also something special
happens to the characteristic class defining the structure, namely to
the second Pontrjagin class $p_2$. For String bundles, $p_2$ is 
divisible by 6, so that $\frac{1}{3}Q_2$ should be used as a generator
instead of simply $Q_2$ ($=\frac{1}{2}p_2$ when $Q_1=0$). 
Then the first generator for the cohomology of BString will be 
$\mathcal{Q}_1:=\frac{1}{3}Q_2=\frac{1}{6}p_2$, and 
\(
H^*(B{\rm String};\Z)\cong \Z[ \mathcal{Q}_1, \mathcal{Q}_2, \cdots] \oplus 
\mathcal{T} \oplus  \mathcal{T}'\;.
\)
We will not attempt to determine the higher generators $\mathcal{Q}_i
\in H^{8i}(B{\rm String};\Z)$, $i \geq 2$, nor
the 2-torsion $\mathcal{T}$ and the 3-torsion $\mathcal{T}'$, as we 
are dealing with relatively low dimensions where the only relevant generator
is $\mathcal{Q}_1$. 


\vspace{3mm}
We can characterize the one-loop term $I_8
=\frac{1}{48}(p_2-\lambda^2)$ 
\cite{DLM} \cite{VW} (which is part of \eqref{dualc})
in the case of a String structure using 
String characteristic classes $\mathcal{Q}_i$. 
We did a similar task in 
\cite{KSpin}, where we wrote the one-loop term in the Spin 
case in terms of the Spin characteristic classes $Q_i$. 
We have: {\it
The one-loop term in M-theory on a String manifold is $\frac{1}{8}\mathcal{Q}_1$}.
We will study the significance of this elsewhere.

%

\section{Relating the Structures}
\label{relate}
The structures we have defined in this paper are related. 
We can find interrelations among them using standard 
arguments. We have already indicated a few such relations in previous sections, and 
other relations can be easily deduced.
For example,  
consider a twisted String structure, given by the obstruction 
$
\lambda + \alpha=0$.
Applying the Steendrod square $Sq^3$ gives
$
W_7 + Sq^3 \alpha=0$,
which is the vanishing of the obstruction to having a String${}^{K(\Z, 3)}$-structure.  

\vspace{3mm}
Using this type of reasoning, we can relate other structures. We can 
also talk about further new higher structures (albeit without immediate
physical applications). Taking a twisted Fivebrane 
structure defined via $\frac{1}{6}p_2+ \beta=0$, 
and applying the Steenrod square $Sq^7$ gives 
\(
Sq^7\left( \frac{1}{6}p_2+ \beta \right)=Sq^7(\mathcal{Q}_1+ \beta)=
Sq^7\mathcal{Q}_1 + Sq^7 \beta\;,
\)
which might be considered as an obstruction to defining 
a {\it twisted Fivebrane${}^{K(\Z,7)}$-structure}. 

\vspace{3mm}
This paper has only achieved a first step in uncovering
the structures discussed, and there 
obviously remains a lot of work to study them systematically further. 
In addition, we hope that geometric and possibly even analytical
descriptions of these structures will be possible in the future.

\vspace{1cm}
{\bf \large Acknowledgements}

\vspace{2mm}
The author would like to thank
the Max Planck Institute for Mathematics in
Bonn for support and for the inspiring atmosphere  
during the writing of this paper in summer 2010.
The author is grateful to 
Jim Stasheff for useful comments on the 
draft.


\end{document}